\documentclass{article}
\usepackage[utf8]{inputenc}
\usepackage{authblk}

\title {Do gravitational waves confirm Hawking's area law?}
\author{Galina Weinstein}
\affil{\normalsize The Department of Philosophy, University of Haifa, Haifa, the Interdisciplinary Center (IDC), Herzliya, Israel.} 

\begin{document}

\maketitle

\begin{abstract} 

Recently an experiment has been performed for the  purpose  of  "testing the area Law with GW150914 data". As the experimenters put it, the experiment presents "observational confirmation" of Hawking’s area law based on the GW150914 data. It is the purpose of this paper to philosophically examine the test of the area law and to show that the area law is not confirmable yet is falsifiable. Accordingly, the GW150914 data do not confirm Hawking's area law. What has been tested with positive results was the hypothesis $A_3 > A_1 + A_2$, where $A_3=$ GW150914 remnant and $A_1 + A_2=$ GW150914 merger. But this single instance does not provide observational confirmation of Hawking's area law.       
\end{abstract}

\section{Introduction}

In 2016, Stephen Hawking commented on the first observation of gravitational waves, the GW150914 gravitational wave event. The gravitational wave signals were observed by the LIGO's two detectors in Hanford and Livingston on 14/09/2015.
Hawking said: "This discovery is the first detection of a black hole binary system and the first observation of black holes merging. The observed properties of this system is consistent with predictions about black holes that I made in 1970 here in Cambridge. The area of the final black hole is greater than the sum of the areas of the initial black holes as predicted by my black holes area theorem. The properties of these black holes are also consistent with the black hole no-hair theorem, which says that they should be uniquely characterized by their mass and spin" \cite{Hawking 2016}. 
After Hawking's death, experiments were performed for the purpose of testing Hawking's theorems.

The goal of this paper is not to reconstruct a historical picture of the detection of gravitational waves. This has been done in great detail both historically and technically by other writers. See for instance \cite{Kennefick} and \cite{Thorne}. 
The primary aim of this paper is to analyse from a philosophical point of view the tests of the no-hair theorem and the area law. 

I begin the paper by describing in Section \ref{tests1} the tests performed by the LIGO Scientific and Virgo Collaborations. It was demonstrated by the collaborations that the Kerr black hole hypothesis is consistent with the GW150914 data. I then discuss in Section \ref{tests}, the test of the no-hair theorem with the GW150914 data. Next, Section \ref{area2} presents the test of the area theorem using the GW150914 data. In Section \ref{area2}, I include a discussion of the contradiction between the area law and Hawking radiation. The paper then proceeds in Section \ref{confirmation} to a discussion of confirmation theory and degree of confirmation. 
Maximiliano Isi and colleagues who have performed the test write in their paper: "We present observational confirmation of Hawking’s black-hole area theorem based on data from GW150914, finding agreement with the prediction [...] when we model the ringdown" phase \cite{Isi2}, p. 1. It is tempting to agree with this statement. I will explain the precise sense in which hypotheses may be confirmable and falsifiable and then show that what has been achieved was not confirmation of Hawking's area law.   

\section{Detection of gravitational waves} \label{Detection}

\subsection{Testing the Kerr black hole hypothesis} \label{tests1}

Two stellar-mass black holes that merge, form a single distorted compact object that gradually settles to a final stationary form. Gravitational waves are emitted throughout the entire process, at each moment carrying information about the evolving compact object. There are three stages in the coalescence of the two black holes \cite{Weinstein}, pp. 18-19:
\vspace{1mm} 

1) \emph{Inspiral}: a long phase in which the two black holes are still quite far one from each other but are slowly orbiting one another in a quasi-circular orbit. Since we are dealing with weak gravitational fields, this phase allows an analytic description using the post-Newtonian expansion (approximation). During this phase, the amplitude and frequency of the gravitational waves are increasing.   

2) \emph{Merger}: the distance between the two black holes gets smaller, the orbit gradually shrinks and the two bodies finally merge. Numerical relativity (numerical simulations) is required. That is because the gravitational field is so strong and also time-dependent. During this phase, the amplitude of the gravitational waves gradually increases until it reaches a maximum.   

3) \emph{Ringdown}: the newly created object releases its final gravitational wave signals away into endless space. It wobbles and oscillates during which it rings like a bell with characteristic frequencies and damping times determined entirely by the mass and spin of the black hole. This stage is treated analytically using perturbation theory. 
A linearly perturbed Kerr black hole emits gravitational waves in the form of exponential damped sine waves, with specific frequencies and decay rates determined exclusively by the black hole’s mass and spin. 

The ringdown phase consists of a superposition of damped sinusoids, quasi-normal modes. Each quasi-normal mode has a characteristic complex angular frequency: the real part is the angular frequency and the imaginary part is the inverse of the damping time. These modes are distinguished by their longitudinal and azimuthal indices, $l$ and $m$ respectively, as well as by their overtone number $n$. Ringdown overtones are the quasi-normal modes with the fastest decay rates. Each mode has a particular frequency and decay rate which are functions of the Kerr black hole parameter spin and total mass of the black hole that is being perturbed. Numerical simulations have demonstrated that the fundamental mode $l = m = 2$ and $n=0$ seems to dominate the ringdown signal \cite{Buonanno}, p. 124018-18, p. 124018-21.
\vspace{1mm} 

In 2016, the LIGO Scientific and Virgo Collaborations announced the first joint detection of gravitational waves (GW150914) with the LIGO detectors. In 2018, they reported of several quantitative tests made on the gravitational-wave data from the detection of GW150914. The purpose of the tests was to show that the Kerr black hole hypothesis (two black holes form a remnant Kerr black hole) is consistent with the data. The result of the tests indicated that the entire GW150914 inspiral-merger-ringdown (IMR) waveform does not deviate from the predictions of a binary black-hole merger in classical general relativity \cite{LIGO}, pp. 221101-3-221101-5.  

It was found that the detected gravitational-wave signal increases in frequency and amplitude in about eight cycles from $35$ to $150$ Hz, where it reaches a peak amplitude. After a time around $0.42 s$, the amplitude drops rapidly, and the frequency appears to stabilize. After the peak gravitational wave amplitude is reached, the signal makes one to two additional cycles, continuing to rise in frequency until reaching about $250$ Hz, while dropping sharply in amplitude. 

According to the LIGO Scientific and Virgo Collaborations, the most plausible explanation for this empirical evolution is gravitational-waves emission from two orbiting masses in the inspiral and merger phases. The drop in amplitude is in accord with a Kerr black hole. 

The LIGO Scientific and Virgo Collaborations concluded that the signal in the data was fully consistent with the final object being a Kerr black hole with a mass $M_{BH_3}=65M\odot$ and a dimensionless spin parameter $a_3 = 0.7$. For a Kerr black hole, the ringdown is expected to have a damping time roughly equal to the period of oscillation (inversely equal to the frequency). \vspace{1mm} 
For a black hole with spin $a_3 = 0.7$, a ringdown frequency of $\approx 260$ Hz$\left(\frac{65M\odot}{M}\right)$ and a damping time of $4ms\left( \frac{M}{65M\odot}\right)$ were calculated. 
\vspace{1mm} 
It was then written: "Such a final mass [$M_{BH_3}=65M\odot$] is consistent with the merger of two black holes of $\sim 35M \odot$ each, after accounting for the energy emitted as gravitational waves" \cite{LIGO2}, pp. 2-3, pp. 12-13; \cite{Weinstein}, pp. 19-20.

More specifically, the following parameters were extracted for the masses and spins of the two progenitor black holes: 

\begin{equation} \label{equation 11}
M_{BH_1}= 36.2^{+5.2}_{-3.8}M \odot, a_1=0.3^{+0.5}_{-0.3},  M_{BH_2}=29.1^{+3.7}_{-4.4} M \odot, a_2=0.5^{+0.5}_{-0.4},    
\end{equation}

\noindent and the measured mass and spin for the remnant that is produced by the coalescence of the binary is \cite{LIGO3}, p. 241102-1, p. 241102-6-241102-7:

\begin{equation} \label{equation 12}
M_{BH_3}=62.3^{+3.7}_{-3.1} M\odot, a_3 = 0.68^{+0.05}_{-0.06},    
\end{equation}

\noindent The above masses are estimated in the source frame and not in the detector frame. Two additional parameters are the energy radiated by the inspiral and merger of the two black holes:

\begin{equation}
E = 3 \pm 0.5 M\odot c^2,
\end{equation}

\noindent and the redshift of the source: 

\begin{equation} \label{equation 15}
z = 0.09 ^{+0.03}_{-0.04}.    
\end{equation}

The LIGO Scientific and Virgo Collaborations explain that the initial behavior of the gravitational-wave signal "cannot be due to a perturbed system returning back to stable equilibrium, since oscillations around equilibrium are generically characterized by roughly constant frequencies and decaying amplitudes". On the other hand, the GW150914 data "demonstrate very different behavior. During the period when the gravitational wave frequency and amplitude are increasing, orbital motion of the two bodies is the only plausible explanation: there, the only 'damping forces' are provided by gravitational wave emission, which brings the orbiting bodies closer (an 'inspiral'), increasing the orbital frequency and amplifying
the gravitational wave energy output from the system" \cite{LIGO2}, p. 3. 
It is concluded: "our inspiral–merger–ringdown test shows no evidence of discrepancies with the predictions of GR" \cite{LIGO}, p. 221101-5. 

The LIGO Scientific and Virgo Collaborations further performed a test that allowed "for possible violations of GR" and concluded that the value of parameters was usually found to represent GR 
\cite{LIGO}, p. 221101-8, pp. 221101-10-11; \cite{Weinstein}, p. 20. 

\subsection{Testing the no-hair theorem} \label{tests}

Astrophysicists have tested the black hole ring at the correct frequencies and damping times, and set themselves to test the hypothesis that states that the two black holes form a remnant Kerr black hole with mass $M_{{BH}}$ and spin $a$. 

Already in 2004, a team of astrophysicists had asked: “can gravitational wave observations provide a test of one of the fundamental predictions of general relativity: the no-hair theorem?” They suggested “a definitive test of the hypothesis that observations of damped, sinusoidal gravitational waves originate from a black hole or, alternatively, that nature respects the general relativistic no-hair theorem” \cite{Dreyer}, p. 787.  

In 2017, three astrophysicists belonging to the LIGO Scientific and Virgo Collaborations, Eric Thrane, Paul Lasky and Yuri Levin, wrote that “The recent detections of gravitational waves from stellar-mass black hole mergers would seem to suggest that a test of the no-hair theorem might be around the corner”. A method was provided "for testing the no-hair theorem using only data from after the remnant black hole has settled into a perturbative state”. 

The post-merger remnant must be allowed to settle into a perturbative, Kerr-like state and this means that the ringdown frequencies and damping times depend only on the mass and spin of the newly created black hole. In this way, the no-hair theorem places stringent requirements on the asymptotic behavior of perturbed black holes. The no-hair theorem concerns itself with linear perturbations. But it was found that at no point in time is the post-merger waveform precisely described by black hole perturbation theory. That is because, there is always a contribution, however small, left over from the merger. Furthermore, the ringdown signal becomes weaker as it settles into a Kerr black hole. 

The authors were confronted with the following problem: they could either obtain higher signal-to-noise (SNR) ratio by adding louder signals from the merger phase (the main gravitational wave signal) or, wait for the remnant black hole to settle to the perturbative state where the no-hair theorem applies. They chose the latter option because as louder signals were added, it was no longer clear whether this improved the knowledge of the ringdown frequency and damping times. So, they arrived at the conclusion that it may be possible to test the no-hair theorem but the observed behavior could either be attributed to a numerical relativity artifact, or to a residual non-linearity of the signal from the merger decaying on a timescale comparable to the linear ringdown signal \cite{Thrane}, pp. 102004-1-2, p. 102004-5.

Matthew Giesler, Mark Scheel and Saul Teukolky have been working for quite some time on the LIGO data and gravitational waves. In May 2019, they submitted a paper with Maximiliano Isi and Will Farr from the LIGO Scientific and Virgo Collaborations in which they analyzed the ringdown gravitational-wave data. They thought that a gestalt shift was needed. It does not mean that Thrane, Lasky and Levin's reservations are not sound. It only means that one team thinks they have solved the problem whilst the other probably holds an opposite view.

Recall that the ringdown phase consists of a superposition of quasi-normal modes. Black hole perturbation theory was used to infer the remnant parameters - mass and spin - from the frequency and damping times of its quasi-normal modes, as imprinted on the later portion of the GW150914 signal. The chirp has a peak amplitude at a peak frequency of $150$ Hz, beyond which the amplitude gradually decreases. Giesler  et  al. started at the peak of the signal and used data starting at different times after the peak frequency of $150$ Hz. "Rather than nonlinearities, times around the peak are
dominated by ringdown overtones — the quasinormal modes with the fastest decay rates, but also the highest amplitudes near the waveform peak" \cite{Isi}, p. 111102-1.

They showed that including enough overtones of the fundamental mode $l = 2$ allows obtaining higher SNR. 
A GW150914-like signal was first studied.\footnote{A simulated output of a LIGO-like detector was studied in response to the same gravitational waves as the ones from the asymptotic remnant. That is, the $l = m = 2$ mode of the signal was injected into simulated Gaussian noise corresponding to the sensitivity of Advanced LIGO in its design configuration \cite{Giesler}, p. 041060-8, p. 041060-10.} It was found that the inclusion of enough overtones associated with the $l = m = 2$ mode “provides a high-accuracy description of the ringdown”, where the high SNR “can be exploited to significantly reduce the uncertainty in the extracted remnant properties” \cite{Giesler}, p. 041060-2. 

Previously, say astrophysicists, overtones had been believed to be too faint to be detected and it was believed that one had to wait at least ten years to test the no-hair theorem. But the team shows that although the overtones decay very quickly, the first overtones of GW150914 should be loud enough to be detected.  
Measuring two quasi-normal modes from gravitational-wave observations is tantamount to measuring the ringdown spectrum. The ringdown spectrum is "a fingerprint that identifies a Kerr black hole" and the measurements of this spectrum "has been
called black-hole spectroscopy" \cite{Isi}, p. 111102-1.

In the spring of 2019, the team, with Isi being the leading author, published a paper, “Testing the No-Hair Theorem with GW150914” \cite{Isi}. They were “Assuming the remnant is a Kerr black hole”. That is, “assuming first that quasinormal modes are as predicted for a Kerr black hole within general relativity”, they checked this assumption \cite{Isi}, p. 111102-2. In other words, the hypothesis that quasinormal modes are as predicted for a Kerr black hole was tested. Recall that the LIGO Scientific and Virgo Collaborations had already tested this hypothesis, see Section \ref{tests1} \cite{LIGO}, p. 221101-5, p. 221101-10; \cite{Weinstein}, pp. 20-21. Now Isi et al. have performed another test. 

Assuming the remnant is a Kerr black hole, using the fundamental mode $l = m = 2$ and its first overtone $n=0$, the GW150914 ringdown was analyzed. Making use of overtones, information was extracted about the GW150914 remnant using only post-inspiral data, starting at the peak of the signal. This analysis led to the detector-frame mass $M_f$ and spin $a_f$ of the remnant:

\begin{equation} \label{equation 14}
M_f = 68 \pm 7 M \odot,   a_f = 0.63 \pm 0.16.    
\end{equation}

\noindent $M_f=M_S(1+z)$, where $M_S$ is the mass measured in the source frame and $z$ is given by equation (\ref{equation 15}).

These are ringdown-only measurements. The ringdown-only measurements of the remnant mass and spin magnitude were then compared to those obtained from the analysis of the entire GW150914 inspiral-merger-ringdown (IMR) waveform. It was concluded that evidence of the mode $l = m = 2$ and at least one overtone was found and, a $90 \%$-credible measurement of the remnant mass and spin was obtained in agreement with that inferred from the full waveform. 

Measuring the frequencies of the fundamental and first overtone, information was extracted about the GW150914 remnant and "their consistency with the Kerr hypothesis" was established. It was concluded "that GW150914 did result in a Kerr black hole as predicted by general relativity, and that the postmerger signal is in agreement with the no-hair theorem". That is, the GW150914 merger "produced a Kerr black hole as described by general relativity" \cite{Isi}, pp. 111102-1-3, p. 111102-5. 

\subsection{Testing the area law} \label{area2}

In 1971 Stephen Hawking considered the following situation. Imagine there are initially two black holes a considerable distance apart. The black holes are assumed to have formed at some earlier time as a result of either gravitational collapses or, the amalgamation of smaller black holes. According to the Israel-Carter uniqueness theorems, the Kerr metric is proved to be the unique stationary, asymptotically flat, vacuum solution with an event horizon. By the said uniqueness theorems the two black holes merge to form a single black hole which settles down to a Kerr black hole with mass $M_{{BH}}$ and spin $a$. The area of the event horizon of the newly formed black hole is given by:

\begin{equation} \label{equation2}
A_K=8 \pi \left(\frac{GM_{{BH}}}{c^2}\right)^2 \left(1+ \sqrt{1-a^2} \right),
\end{equation}

\noindent Hawking shows that the area $A_K$ of the event horizon of the resulting black hole is greater than the sum of the areas of the event horizons around the original black holes:

\begin{equation} \label{equation10}
A_K\geq(A_1+A_2). 
\end{equation}

\noindent Hawking writes equation (\ref{equation2}) in the following form:

\begin{equation} \label{equation7}
M_{{BH_3}}^2 \left(1+ \sqrt{1-a_3^2} \right)> M_{{BH_2}}^2 \left(1+ \sqrt{1-a_2^2} \right)+M_{{BH_1}}^2 \left(1+ \sqrt{1-a_1^2} \right),
\end{equation}

\noindent and says: "By the conservation law for asymptotically flat
space, the energy emitted in gravitational or other forms of radiation is:
\vspace{1mm} 

\noindent $M_{BH_1} + M_{BH_2} - M_{BH_3}$. 
\vspace{1mm} 
\noindent The efficiency:
\vspace{1mm} 

\noindent $\epsilon = \left(M_{BH_1} + M_{BH_2}- M_{BH_3}\right)/ \left(M_{BH_1} + M_{BH_2} - M_{BH_3}\right)$ 

\vspace{1mm} 
\noindent is limited by Eq. (\ref{equation7})". 

\noindent Hawking gives the upper limits of the efficiency $\epsilon$ and stresses that the actual efficiency might be much less. 

\noindent The highest limit on $\epsilon$ is $\frac{1}{2}$ which occurs if two maximally spinning equal-mass black holes $M_{BH_1} = M_{BH_2} = a_1 = a_2$ create a non-spinning Schwarzschild black hole: $a_3=0$. Equation (\ref{equation7}) then becomes:

\begin{equation} \label{equation8}
M_{{BH_3}}^2 > M_{{BH}_{1,2}}^2 \left(1+ \sqrt{1-a_{1,2}^2} \right).
\end{equation}

\noindent For two non-spinning black holes $a_1 = a_2=0$ that merge together to form a spinning black hole:

\begin{equation} \label{equation9}
M_{BH_3}^2 \left(1+ \sqrt{1-a_3^2} \right)> 2M_{{BH_2}}^2 +2M_{{BH_1}}^2,
\end{equation}

\noindent and $\epsilon < 1- \frac{1}{\sqrt{2}}$ and so $\epsilon < \frac{1}{2}$ \cite{Hawking 1971}, p. 1345.
\vspace{1mm} 

In 1973 James Maxwell Bardeen, Carter and Hawking formulated the four laws of black hole mechanics, the second law of which is Hawking’s 1971 area law \cite{BCH}, pp. 167-168:
"The area $A$ of the event horizon of each black hole does not decrease with time $\delta A \geq 0$.

\noindent If two black holes coalesce, the area of the final event horizon is greater than the sum of the areas of the initial horizons, i.e.

\noindent $A_3 > A_1 + A_2$".
\vspace{1mm} 

Three astrophysicists from CERN (Gian Giudice, Matthew McCullough and Alfredo Urbano) have considered the detection of GW150914 as an observational test of Hawking's area law. 
Inserting equations (\ref{equation 11}) and (\ref{equation 12}) into equation (\ref{equation7}), they show that the area law is consistent with the data. Equations (\ref{equation 11}), (\ref{equation 12}) and (\ref{equation2}) give:

\begin{equation}
A_1 + A_2 = \left(2.2 \pm 0.2 \right) \times 10^5 km^2,
A_K = \left(3.8 \pm 1.3 \right) \times 10^5 km^2.
\end{equation}

\noindent According to the team, what equations (\ref{equation 11}) and (\ref{equation 12}) show us is that Hawking's law (\ref{equation10}) is consistent with the GW150914 data. 
"This result shows that the observation of GW150914 has verified equation (\ref{equation7}), hence Hawking’s area theorem, well within errors", say the three astrophysicists \cite{GMU}, p. 32.

After testing the no-hair theorem with the GW150914 data, Isi et al. tested the area law. 
According to equation (\ref{equation2}), for two well-separated inspiraling progenitor black holes, the total horizon area is:

\begin{equation} \label{equation 10}
A_0 \equiv A \left(M_1, a_1 \right) + A \left(M_2, a_2 \right),     
\end{equation}

\noindent where $M_{1,2}$ and $a_{1,2}$ are the masses and spins of the
two black holes.

The merger produces a remnant black hole with mass and spin $M_f$ and $a_f$ whose horizon area is:

\begin{equation} \label{equation 13}
A_f \equiv A \left(M_f, a_f \right),    
\end{equation}

\noindent where $A_f$ is defined by equation (\ref{equation2}), and $a_f=a$.

\noindent Isi et al. proceed to extract $A_0$ and $A_f$ from the GW150914 signal in order to compute the change in the total area \cite{Isi2}, pp. 2-3:

\begin{equation}
\Delta A \equiv A_f - A_0.    
\end{equation}

For the ringdown data, the measurements of $M_f$ and $a_f$ are given by equation (\ref{equation 14}). $M_{1,2}$ and
$a_{1,2}$ are estimated using the Python \texttt{NRSur7dq4} numerical relativity waveform surrogate model. This gives an
accurate representation of the signal up to the peak at frequency $150$ Hz. The final result is that the measurements favor $\Delta A \geq0$ in agreement with Hawking’s area law. Isi et al. assert that the area law is confirmed "with $97 \%$ credibility if relying on the overtone" [the overtones of the mode $l = m = 2$, i.e. on equation (\ref{equation 14})] or $95 \%$ if not" \cite{Isi2}, p. 2. 
It is concluded: "We present observational confirmation of Hawking’s black-hole area theorem based on data from GW150914, finding agreement with the prediction with $97 \%$ ($95 \%$) probability when we model the ringdown including (excluding) overtones of the quadrupolar mode" \cite{Isi2}, p. 1.

Isi et al. explain that Hawking's area law is "a fundamental consequence of general relativity (GR) and the cosmic censorship hypothesis" \cite{Isi2}, p. 1.

Indeed, naked singularities refute Hawking’s area law. Let us see why this happens. If the horizon of the black hole disappears, the area of the horizon gets smaller and smaller and the area law is violated. The solutions of the field equations of general relativity which describe naked singularities not hidden by an event horizon cause severe problems to the Isreal-Carter uniqueness theorems. The reason is that only if all singularities are surrounded by event horizons, the Kerr black hole is a unique solution. So, in 1969 and 1976 Roger Penrose and Hawking phrased the censorship hypothesis, which says that physics censors naked singularities by always enshrouding them with a horizon \cite{Penrose},  p. 1160, p. 1162; \cite{Hawking 1976}, p. 2461. 

First, in 1969 Penrose introduced the idea of a “cosmic censor” who forbids the appearance of naked singularities, clothing each one in an event horizon. He added that it is not known whether naked singularities would ever arise in a collapse which starts off from a non-singular initial state \cite{Penrose}, p. 1160. 

The point is that Hawking radiation and evaporation refute Hawking's area law because the mass of the black hole decreases and consequently the area of the black hole gets smaller and smaller as the black hole radiates Hawking radiation.
So, in 1975 Hawking formulated \emph{“The ‘cosmic censorship’ hypothesis: Nature abhors a naked singularity”}, namely, any singularities which are developed from gravitational collapse will be hidden from the view of observers at infinity by an event horizon. Hawking pointed out that evaporation violates the classical censorship hypothesis. He explained that if one tries to describe the process of a black hole losing mass and eventually disappearing and evaporating by a classical space-time metric, then “there is inevitably a naked singularity when the black hole disappears. Even if the black hole does not evaporate completely one can regard the emitted particles as having come from the singularity inside the black hole and having tunneled out through the event horizon on spacelike trajectories” \cite{Hawking 1976}, p. 2461. 

As the black hole radiates Hawking radiation, the area of the black hole gets smaller and smaller. Hence, Hawking radiation violates the area law but does not contradict the generalized second law of thermodynamics (GSL). Although Hawking radiation causes the surface area $A_K$ to steadily decrease, the entropy of the black hole $S_{BH}$ plus the entropy of the matter outside of the black hole $S_M$ will be larger than before. Eventually the mean sum of the change of entropies $dSg=dS_{BH}+dS_M>0$ will increase.
The GSL makes a stronger statement than Hawking’s area law. The area law requires only that the area not decrease. But Jacob Bekenstein writes that "in the astonishing quantum process of spontaneous radiation by a Kerr black hole discovered by Hawking, the area theorem is flagrantly violated, but the increase in exterior entropy due to the radiation is expected to suffice to uphold the GSL” \cite{Bekenstein 1975}, p. 3077, p. 3079.

\section{Confirmation of hypotheses} \label{confirmation}

"How do we find confirmation of a law?" asked Rudolf Carnap and answered: "If we have observed a great many positive instances and no negative instance, we say that the confirmation is strong. How strong it is and whether the strength can be expressed numerically is still a controversial question in the philosophy of science". So, can the degree of confirmation of a law be expressed in a quantitative form? instead of saying that one law is "well founded" whilst the other "rests on flimsy evidence", Carnap suggests that "we might say that the first law has a $0.8$ degree of confirmation, whereas the degree of confirmation for the second law is only 0.2". Hence, what Carnap calls "degree of confirmation" is "identical with logical probability". Unlike statistical probability, logical probability according to Carnap means that "if the evidence is so strong that the hypothesis follows logically from it - is logically implied by it - we have one extreme case in which the probability is 1". By the same token, "if the negation of a hypothesis is logically implied by the evidence, the logical probability of the hypothesis is 0". For all the cases in between, neither the hypothesis nor its negation can be deduced from the evidence and we are justified in assigning a numerical value to the probability \cite{Carnap}, pp. 167-168.  

It should be stressed that if by confirmation is meant a complete, a definitive and final establishment of truth, then a hypothesis can never be confirmed. There is no complete confirmation possible but only a process of gradually increasing confirmation. We can only confirm a hypothesis more and more but it is not confirmable because the number of instances to which the law refers is infinite and therefore can never be exhausted by our observations which are always finite in number. Although we cannot verify the law, we can test it by testing its single instances i.e. the particular hypotheses which we derive from the law and from other hypotheses established previously. If in the continued series of such testing experiments, no negative instance is found but the number of positive instances increases then our confidence in the law will grow. We say that the degree of confirmation, after a few observations have been made, will be so high that we practically cannot help accepting the hypothesis. But there always remains still the theoretical possibility of denying it. Consequently, says Carnap, every hypothesis is a probability-hypothesis \cite{Carnap2}, p. 420, pp. 425-426.\footnote{Some philosophers would later define the degree of probability as a quantitative concept and interpret the degree of confirmation of a hypothesis as the degree of probability in the strict sense which this concept has in the calculus of probability.}

Accordingly, we cannot verify the no-hair theorem that says "all black holes are the same" because the number of black holes in the universe is infinite. We can, however, test the no-hair theorem by testing its single instances. Astrophysicists have tested the particular hypothesis that states that two black holes form a remnant Kerr black hole with mass $M_{{BH}}$ and spin $a$. They found that the two progenitor black holes GW150914 produced a GW150914 Kerr black hole. 
As seen in Section \ref{tests}, this is exactly the conclusion arrived at by Isi et al. with respect to the no-hair theorem. They write in the conclusion of their first paper: "The agreement between postinspiral measurements with two different sets of modes supports the hypothesis that GW150914 produced a Kerr black hole as described by general relativity. Moreover, we constrain deviations away from the no-hair spectrum by allowing the overtone frequency and damping time to vary freely. This is equivalent to independently measuring the frequencies of the fundamental and first overtone, and establishing their consistency with the Kerr hypothesis" \cite{Isi}, p. 111102-5.

Like Carnap, Carl Hempel adopts the term "degree of confirmation" instead of probability "because the latter is used in science in a definite technical sense involving reference to the relative frequency of the occurrence of a given event in a sequence, and it is at least an open question whether the degree of confirmation of a hypothesis can generally be defined as a probability in this statistical sense". 

Hempel seeks to characterize in precise and general terms the conditions under which a body of evidence can be said to confirm, or to disconfirm, an empirical hypothesis. According to Hempel, the criteria of empirical confirmation should be formal, objective and should contain no reference to the subject-matter of the hypothesis and the evidence in question. With this goal in mind, he constructs a definition of confirmation and 
formulates some basic principles of what he calls "the logic of confirmation". He argues that "an adequately defined concept of confirmation should satisfy" these conditions \cite{Hempel1}, pp. 6-9. 

Hempel is most known for "the paradox of confirmation" ("all ravens are black"). Consider first the sentence: "Twins always resemble each other". A confirming instance consists of two persons who are twins and resemble each other; twins who do not resemble each other would disconfirm the hypothesis; and any two persons not twins - no matter whether they resemble each other or not - would constitute irrelevant evidence. Only observations of twins constitute relevant evidence. This is the most tacit interpretation of the concept of confirmation. Hempel shows that this criterion (which he calls Nicod's criterion) suffers from serious shortcomings, the most serious of which is the following:     
Consider the two sentences, $S_1$: "all ravens are black" and $S_2$: "whatever is not black is not a raven". Let $a$ and $b$ be two objects such that $a$ is a black raven and $b$ is a white chair (neither a raven nor black). $S_1$ and $S_2$ are logically equivalent. They are different formulations of the same hypothesis. The gist of the paradox is that $a$ would confirm $S_1$ and $b$ would confirm $S_2$. But since $S_1$ and $S_2$ are logically equivalent, $b$ would also confirm $S_1$ \cite{Hempel1}, pp. 10-11. 

So, for instance, the no-hair theorem states that "all black holes are the same". They all have no-hair. Let us call this sentence $S_1$. The sentence $S_2$ "whatever is not the same is not a black hole" is logically equivalent to $S_1$. Let $a$ and $b$ be two objects such that $a$ is a black hole with no-hair and $b$ are black and white chairs. $a$ would confirm $S_1$ and $b$ would confirm $S_2$. Again, since $S_1$ and $S_2$ are logically equivalent, $b$ would also confirm $S_1$. 
Thus, observing more instances of hairless black holes would not confirm the no-hair theorem.

According to Hempel, no finite amount of experimental evidence can conclusively confirm a general or theoretical law of nature. But a finite set of relevant data may well be "in accord with", i.e. be consistent with an empirical hypothesis. The acceptance of a general hypothesis on the basis of a sufficient body of evidence will as a rule be tentative. It will hold only with the proviso that if new and unfavourable evidence should turn up the hypothesis will be abandoned again. It is always possible that new evidence will be obtained which dis-confirms the hypothesis \cite{Hempel1}, p. 2; \cite{Hempel2}, p. 116. 

If as usually assumed the universe contains an infinite number of objects, then the following general hypothesis, "There are Kerr black holes", is verified by the evidence (example): "Object $A$ is a Kerr black hole". But no finite observational example can contradict and thus falsify the hypothesis. Conversely, a purely universal hypothesis such as "all black holes are Kerr black holes" is falsifiable but not verifiable, i.e. confirmable, for an infinite universe of discourse. The above general hypothesis is completely falsified by the observation of a Schwarzschild black hole. But no finite observations can entail and thus confirm the hypothesis in question. Observational examples can never confirm the general hypothesis "all black holes are Kerr black holes". But the more observational examples of Kerr black holes we collect, the more confident we are in the general hypothesis  \cite{Hempel2}, p. 113.

So for instance, the hypothesis "there are Kerr black holes" can be verified by the evidence: "the GW150914 merger created a GW150914 remnant which is a Kerr black hole”.
As seen in Section \ref{tests}, measuring the frequencies of the fundamental and first overtone, information was extracted about the GW150914 remnant and it was concluded that "the  GW150914 merger ”produced a Kerr black hole as described by general relativity” \cite{Isi}, pp.111102-1-3, p. 111102-5. 

On the other hand, the general hypothesis, "The area $A$ of the event horizon of each black hole does not decrease with time" (\cite{BCH}, p. 167) is not confirmable. 
What the above hypothesis says is that all black holes in the universe posses the property that the area of their event horizon does not decrease with time. According to the tacit assumption of confirmation, which had been proposed by Nicod, all astrophysicists need to do in order to confirm this law is check that black holes do not refute this law. Of course this is impossible because the universe contains an infinite number of black holes. Moreover, Hempel showed that this method has shortcomings especially with respect to objects other than black holes (the paradox of confirmation). 

So, what astrophysicists do instead is test the area law by testing its single instance: "We infer the premerger area, $A_0$, from the inspiral alone. We infer the postmerger area, $A_f$, from the remnant mass and spin as estimated from an analysis of the ringdown using the fundamental mode and one overtone at the peak, as well as solely the fundamental mode 3 ms after the peak. For the former (latter), we measure $\Delta A / A_0 = 0.52 ^{+0.71}_{-0.47}$ $0.60 ^{+0.82}_{-0.60}$ at $90 \%$ credibility, and find agreement with Hawking’s area theorem with $97 \%$ ($95 \%$) probability" \cite{Isi2}, p. 2. What has been tested with positive results was the hypothesis $A_3 > A_1 + A_2$, where $A_3=$ GW150914 remnant and $A_1 + A_2=$ GW150914 merger. But this single instance does not provide observational confirmation of Hawking's area law.

\section*{Acknowledgement}

This work is supported by ERC advanced grant number 834735.


\begin{thebibliography}{23}

\bibitem[1]{BCH} Bardeen, J. M., Carter, B., Hawking, S. W. (1973). “The Four Laws of Black Hole Mechanics.” \emph{Communications in Mathematical Physics} 31, pp.
161-170.

\bibitem[2]{Bekenstein 1975} Bekenstein, J. D. (1975). “Statistical black-hole thermodynamics.” \emph{Physical Review D}12, pp. 3077-3085.

\bibitem[3]{Buonanno} Buonanno, A., Cook, G. B., and Pretorius, F. (2007). "Inspiral, merger, and ring-down of equal-mass black-hole binaries." \emph{Physical Review D} 75, pp. 124018-1-124018-42.

\bibitem[4]{Carnap2} Carnap R. (1936). “Testability and Meaning.” \emph{Philosophy of Science} 3, pp. 419-471.

\bibitem[5]{Carnap} Carnap R. (1966). “The Confirmation of Laws and Theories.” In Kourany J. A. (ed.), \emph{Scientific Knowledge. Basic Issues in the Philosophy of Science}, 2nd ed. Ca: Wadsworth Publishing Company, 1998, pp. 164-175.

\bibitem[6]{Dreyer} Dreyer, O., Kelly, B., Krishnan, B. Finn, L. S., Garrison, D., and Lopez-Aleman, R. (2014). "Black-hole spectroscopy: testing general relativity through gravitational-wave observations." \emph{Classical and Quantum Gravity} 21, pp. 787–803.

\bibitem[7]{Giesler} Giesler, M., Isi, M., Scheel, M. A., and Teukolsky, S. A. (2019). "Black Hole Ringdown: The Importance of Overtones." \emph{Physical Review X} 9, pp.  041060-1-041060-13.

\bibitem[8]{GMU} Giudice, G. F., McCullough, M. and Urbano, A. (2016). "Hunting for dark particles with gravitational waves." \emph{Journal of Cosmology and Astronautical Physics} 10, pp. 1-43.  

\bibitem[9]{Hawking 1971} Hawking, S. W. (1971). “Gravitational Radiation from Colliding Black Holes.” \emph{Physical Review Letters} 26, pp. 1344-1346.

\bibitem[10]{Hawking 1976} Hawking, S. W. (1976). “Breakdown of predictability in gravitational collapse.” \emph{Physical Review D} 14, pp. 2460-2473.

\bibitem[11]{Hawking 2016} Hawking, S. W. (2016). "Stephen Hawking Congratulates LIGO Team." \emph{Courtesy of BBC News}, Feb 12.

\bibitem[12]{Hempel1} Hempel, C. G. (1945). "Studies in the Logic of Confirmation (I.)." \emph{Mind} 54, pp. 1-26.

\bibitem[13]{Hempel2} Hempel, C. G. (1945). "Studies in the Logic of Confirmation (II.)." \emph{Mind} 54, pp. 97-121.

\bibitem[14]{Isi} Isi, M., Giesler, M., Farr, W. M., Scheel, M. A., and Teukolsky, S. A. (2019). "Testing the No-Hair Theorem with GW150914." \emph{Physical Review Letters} 123, pp. 111102-1-111102-6.

\bibitem[15]{Isi2} Isi, M., Farr, W. M., Giesler, M., Scheel, M. A., and Teukolsky, S. A. (2020). "Testing the black-hole area law with GW150914." \emph{ArXiv} 2012.04486v1[gr-qc], pp. 1-4.

\bibitem[16]{Kennefick} Kennefick, D. (2020). "The Wagers of Science." In Buchwald, J. D (ed.), \emph{Einstein Was Right: The Science and History of Gravitational Waves}, Princeton; Princeton University Press, pp. 62-75.

\bibitem[17]{LIGO3} LIGO Scientific and Virgo Collaborations (2016). "Properties of the Binary Black Hole Merger GW150914." \emph{Physical Review Letters} 116, pp. 241102-1-241102-19.

\bibitem[18]{LIGO} LIGO Scientific and Virgo Collaborations (2018). "Tests of General Relativity with GW150914." \emph{Physical Review Letters} 116, pp. 221101-1-221101-19.

\bibitem[19]{LIGO2} LIGO Scientific and Virgo Collaborations (2017). "The basic physics of the binary black hole merger GW150914." \emph{Annalen der Physik} 529, pp. 1-17.

\bibitem[20]{Penrose} Penrose, R. (1969). “Gravitational Collapse: The Role of General Relativity.” \emph{Rivista del Nuovo Cimento, Numero Speziale} 1, pp. 252-275; \emph{General Relativity and Gravitation} 34, 2002, pp. 1141-1165.

\bibitem[21]{Thorne} Thorne, K. S. (2020). "One Hundred Years of Relativity: From the Big Bang to Black Holes and gravitational Waves." In Buchwald, J. D (ed.), \emph{Einstein Was Right: The Science and History of Gravitational Waves}, Princeton; Princeton University Press, pp. 19-46.

\bibitem[22]{Thrane} Thrane, E. Lasky, P. D., and Levin, Y. (2017)."Challenges for testing the no-hair theorem with current and planned gravitational-wave detectors." \emph{Physical Review D} 96, pp. 102004-1-102004-6.

\bibitem[23]{Weinstein} Weinstein, Galina (2021). "Why Do You Think It is a Black Hole?" \emph{arXiv}:2102.02592 [physics.hist-ph]. 


\end{thebibliography}
\end{document}